\documentclass[letterpaper, 11pt]{article}
\usepackage[margin=0.75in]{geometry}
\usepackage{microtype}
\usepackage{graphicx}
\usepackage{authblk}
\usepackage[hidelinks]{hyperref}
\usepackage{xcolor}
\usepackage{mathpazo}
\usepackage{cite}
\usepackage{amsmath}
\usepackage{mathtools}
\usepackage{cuted}
\usepackage{float}
\usepackage[font={footnotesize, sf}, labelfont=bf, margin=0.05\linewidth]{caption}
\usepackage{lineno}
\usepackage{color}

\title{Be Prospective, Not Retrospective: A Philosophy for Advancing Reproducibility in Modern Biological Research}

\author[]{Griffin Chure}
\affil[]{Department of Biology, Stanford University, Stanford, CA, USA.}
\affil[]{\texttt{griffinchure@gmail.com}}
\date{\today}

\begin{document}
\maketitle
\begin{abstract}
The ubiquity of computation in modern scientific research inflicts new
challenges for reproducibility. While most journals now require code
and data be made available, the standards for organization, annotation, and
validation remain lax, making the data and code often difficult to decipher or practically use. I believe that this is due to the documentation, collation, and validation of code and data only being done in retrospect. In this essay, I reflect on my experience  contending with these challenges and present a philosophy for prioritizing 
reproducibility in modern biological research where balancing computational 
analysis and wet-lab experiments is commonplace. 
Modern tools used in scientific workflows (such as GitHub repositories) lend 
themselves well to this philosophy where reproducibility begins at project \textit{inception},
not \textit{completion}. To that end, I present and provide a programming-language agnostic template architecture
that can be immediately copied and made bespoke to your next paper, 
whether your labwork is wet, dry, or somewhere in between.  
\end{abstract}

\section*{Introduction}

I entered graduate school in the Fall of 2013 determined to become a  
biophysicist even though my undergraduate training was almost entirely focused
on qualitative molecular biology and biochemistry. While I anticipated the long
nights of catching up with concepts from physics and the various mathematical
methods that they required, I had not anticipated how difficult it would be to
juggle my newfound love for computational research with my expertise in wet-lab
biology. I struggled in keeping my physical lab notebook---filled with images of
gels, marginal calculations of dilution factors, and the occasional stain of buffer---in logical sync with the code I would use to process 
microscopy images or explore some aspects of my theoretical work. The code I
wrote was cryptic with sparse documentation, confusing variable
names, and paths to directories that sometimes had never existed. All of this finally came to a head in one night of misery before my PhD candidacy exam. Needing to rerun some piece of analysis from the previous year, I opened my
Python script only to find that it was hard-coded to read data from a folder on my Desktop
that had been deleted several months prior. Frustrated and sleep deprived at the time, I saw this as a personal scientific failure. I can now look back at that night with the certainty it transformed the way I would approach my science for the rest of my PhD and beyond. After my candidacy exam, I vowed I would never again burn myself (nor anybody else) by having my science be unorganized and irreproducible. 

Whether or not science is really experiencing a reproducibility crisis\cite{fanelli2018,baker2016,piper2020}, accessing raw data or code from other scientists is often a difficult endeavor.  This has been recently demonstrated by Gabelica \textit{et al.}
\cite{gabelica2022} who attempted to obtain data listed as ``available upon
reasonable request" from $\approx$ 1800 recent papers in the biosciences. Of
these articles, only $\approx$ 7\% ultimately shared their data, meaning that
93\% of the studies could not pass even the first stage of reproduciblity. The reasons for this low response rate are varied, but are similar to those
published in another recent meta-analysis from Stodden \textit{et al.} \cite{stodden2018}.
While more tightly focused in scope, they had similar issues and were ultimately able to receive
data and code from about 35\% of their $\approx$ 200 queries. Even when data was provided, the authors were
able to reproduce the scientific results from only $\approx$ 60\%. In cases
where data was not shared, the reasons varied from institutional/ethical
restrictions to outright refusal as their ``code was not
written with an eye toward distributing for other people to use." (Ref. \cite{stodden2018}, p. 2585). 
This can create a slew of problems. Trisovic \textit{et al}.\cite{trisovic2022} recently demonstrated that only $\approx$ 25 \% of code released alongside research papers could be run without error. This represents a view of computation held by many scientists; it's an exercise in personal research, never intended to be used by someone else. This pulls me back to that fateful
night in preparing for my candidacy exam. Not only did I write that code without an
eye towards sharing with others, I didn't even write it for my future
self.

Recent years have seen a flurry of excellent papers outlining best practices for
reproducible research, spanning from scientific programming guidelines
\cite{wilson2014a, balaban2021, green2011, lee2018}, to general and specialized data
annotation \cite{rashid2020, falkingham2018}, to instructions for bundling entire
projects as ``reproducible packages" \cite{krafczyk2021} and I encourage the reader 
to give them a look. However, I take a different approach in this essay and give my perspective as a  
practicing biologist who thinks about how to maximize reproducibility 
alongside designing, executing, and analyzing experiments. 

\section*{Data as modern scientific currency}
I view research  as a journey with the generation, manipulation, visualization, and interpretation of data as the overarching themes. Here, I take very  general definition of ``data" to mean ``a collection of qualitative or quantitative facts" such that results from simulations, mathematical analysis, and bench-top experiments are treated equivalently as data-generating processes. While we often remark that the ``data speak for themselves'', this is never truly the case. Not only do you give the data their voices, you give them the language they speak. Reproducibility requires a Rosetta stone such that anyone can perform the translation and come to the same results. 

\begin{figure}[h]
    \centering
    \includegraphics{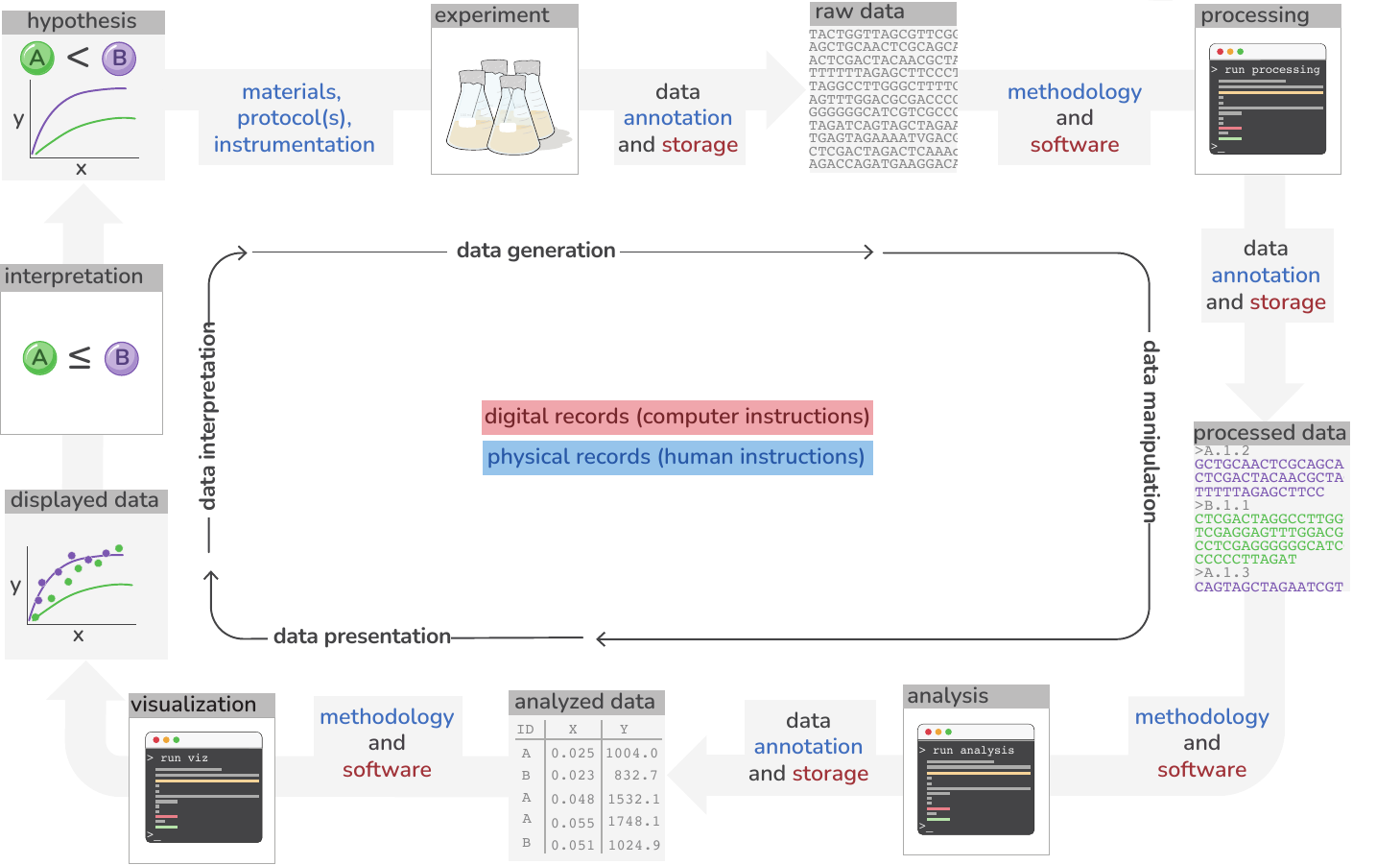}
    \caption{\textbf{The data-centricity of the modern scientific process.} Contemporary scientific research is becoming more centered on data, with scientific findings emerging from successive steps of data generation, manipulation, visualization, and interpretation. Beginning with a hypothesis, experiments (either in the wet- or dry-lab) are carefully designed and appropriately controlled to test their predictions. The raw data is then stored (either physically or digitally) and is later used in downstream analyses, needing to pass through data processing, data analysis, and data visualization steps before conclusions can be drawn and hypotheses can be refined. All of these steps require human-readable documentation (blue text) and computer-readable instructions (red text) in most modern biological research. This diagram shows a workflow for hypothesis-driven research, but can be similarly applied to exploratory or meta-analysis research projects.}
    \label{fig:cycle}
\end{figure}

Consider the ``typical" cycle of science as depicted in Figure \ref{fig:cycle}. Beginning with hypotheses, experiments are designed to thoroughly test and falsify them\footnote{In exploratory research, experiments are designed to properly collect data from which hypotheses will be drawn. In meta-analyses, the ``experiments'' may be collection of data from previously published papers or other resources. In either case, the cycle shown in Figure \ref{fig:cycle} still applies.}, resulting in the generation of new data. These data, whether they come from tangible or computational experiments, often need to be manipulated through processing, cleaning, and analysis pipelines before they can be truly understood. In all cases, these data must be visualized in a way where the experimenter can use their expertise and logical creativity to interpret the results, allowing conclusions to be drawn and the hypothesis to be confirmed, refuted, or refined. In the modern scientific enterprise,  each of these steps require a combination of instructions that are physical and targeted to humans (protocols, observations, notes, etc.) and digital records which are computer-readable (code, instrument settings, accession numbers, etc.). In order for this process to be reproducible, each of these steps must have their instructions meticulously kept and clearly documented. With enough care, these instructions 
come together to serve as your Rosetta stone. 

\section*{Philosophical pillars for reproducibility}
``Making your research reproducible" is easier to say than to do. Through my years of experience in prioritizing reproducibility in my own work, I've found four key principles to be critical to performing my research in a reproducible manner [Figure \ref{fig:principles}(A)]. While the detailed structure or the questions I pose may not be appropriate for your particular project or experiment, the philosophy behind it will likely still apply. This allows you to make a tailor-made reproducible workflow from the ground up in a way that others can follow.

\subsection*{I: Reproducibility begins at project inception, not completion}
The cycle shown in Figure \ref{fig:cycle} produces one scientific finding at a time. As it's now commonplace for modern scientific papers to consist of a surfeit of findings, many different tours of this cycle are often needed. This may make it seem like the only way to explain how to reproduce it is to approach it retrospectively; to wait until the dust settles to list what worked and how it was done while ignoring the failures and bumps along the way. The issue with this approach is that the complications and bumps along the way are often insightful and necessary to understand how and why the data was generated, which is a requirement for reproducible interpretation of data. 

Reproducibility requires intention and effort in structuring the research from the outset. Whenever I start a new project, I ask myself some basic questions: What types of experiments will I undertake? What types of equipment will I need? What types of data will they generate? How ``big'' will the data likely be? Do I need to write my own software to do the analysis?  Even though I never know the exact answers, just having a general idea helps me sketch out what the project structure will be and, more importantly, allows me to identify what physical and digital records I'll probably need to compose. Having these questions and answers be front of mind helps me be intentional in generating these records with an eye towards reproducibility, while retaining the flexibility to adjust them as the research progresses. 

\subsection*{II: Draft your workflow, even though you know it will change}
I have never started a research project where I knew all of the experiments I would do and where all of the twists and turns would take me.  However, for every experiment I \textit{did} do, I had a very good idea of what the workflow would be as I went through the cycle codified in Figure \ref{fig:cycle}. For example, many of my research projects involve measuring the rate of bacterial growth using a 96-well plate reader assay. Even before I run the experiment, I know that I will need to (a) convert the data from an Excel spreadsheet to a  \texttt{csv} file\footnote{Comma Separated Value (\texttt{csv}) files are common in my field of research, though are not necessarily so in other disciplines.}, (b) restrict the measurements to only consider the exponential growth regime, (c) perform some inference of the growth rate using standard tools of regression, and (d) visualize the data along with the regression and its associated uncertainty. Knowing these steps tells me that I will need custom software for cleaning and formatting the data, performing the statistical inference, and generating the 
plots, all of which must read data files. This then generates more questions: How should I name the files? How should they be organized? How will I distinguish between replicates? Sketching out how the data you generate will pass through the steps of manipulation and presentation will help you write your custom code with more intention and, more importantly, will make you assess how to structure your data organization such that analyses can be easily re-run and specific data sets can be easily found.

\subsection*{III: Document as you go along, not just when it's requested}
At some point in your career, you will probably receive an email that reads something like this: ``Do you remember that experiment you performed four years ago? The one where you measured the thing? How did you control for the other thing? Could you send me a detailed protocol so I could use it in my next experiment?" I wager you'll get this email because methods sections are too often approximations of reality---they present an idealistic view of how the average experiment was done, with brevity being heavily prioritized. The devil is in the details and it is enormously helpful to know exactly what was done for a given experiment.

The mutability of memory makes recalling exactly what was done a tricky endeavor. Thus, it's far better to document your experiments and analysis as you go along, rather than just when it's requested by a colleague. I approach each individual experiment as a minimal, reproducible unit. As I go along through the experiment, I take breaks to type\footnote{In my opinion, all research notebooks should be preserved as digital objects and stored on redundant back up systems. I keep my notes as a series of text files backed up locally and stored on version-controlled GitHub repositories. This makes them searchable and shareable in a way that paper notebooks cannot be.} into my notebook exactly what I was doing as I did it, keeping note that even the smallest detail may be an insightful datum for my future self. As an experiment goes on, I (again) ask myself more questions: What materials am I using? Where did I get them? What protocol am I following? How have I deviated from the idealistic protocol during this particular experiment? I have found that answering these questions not only allows me to take extensive notes of what I did on a particular day, it makes me more attentive to detail. 

In the biological sciences, it is routine to perform experiments multiple times to generate enough ``biological replicates" to make statistically satisfying assessments of hypotheses. This means that particular protocols are repeatedly followed with few major changes. To avoid rewriting every step of the protocol for every experiment, I create templates with big, bold \textbf{XXX}'s marking where I will need to add in details that vary across replicates. 

\subsection*{IV: Separate your code by what it does, not when you need it}
The distinction between software and science is rapidly blurring with today's scientist being a technical writer, experimentalist, scholar, and developer simultaneously. It is not uncommon to see entire suites of custom-written software accompanying research papers, even beyond method development. Modern computational tools such as Jupyter Notebooks, where code and prose can be extensively interleaved, are excellent tools for pedagogy\cite{perkel2018, cardoso2019}, but can have significant shortcomings when it comes to reproducibility\cite{pimentel2019, wang2020}. Part of the reason for this poor reproducibility is the desire to fit the entire analysis---from data generation to interpretation---into a single document. Thus, if one wants to tweak aspects of some statistical inference or merely correct a typo in a plot title, the entire script or notebook may need to be rerun.

Separating my code by what it does, and not when I use it, has been remarkably powerful in making my research more reproducible. Let's again consider my typical experiment of measuring bacterial growth rates using a 96-well plate assay. This is an experiment I repeat dozens of times for a single project, meaning that I will have to run the same code over and over again to generate the results. For each experiment, I typically write three different scripts; one that processes and cleans the data, one that performs the statistical inference, and a third which generates the necessary visualizations. As each individual script performs a single step of the scientific cycle, rerunning bits and pieces of my experimental analysis becomes far simpler. 

I try to write all custom software as installable packages, reducing the variability between scripts used for replicate experiments. For example, I may want to write a function that converts Excel-based data to a different file format, as it is a process I must do every time I measure a growth curve. I define a function for this once, storing it in a module of the package specific to data processing and cleaning. I then call this code repeatedly in a script unique to each experimental replicate. Following this approach ensures that the data for every experiment is cleaned the exact same way (a requirement for complete reproducibility), reducing the errors introduced by copy-pasting code from script to script. The most common scientific programming languages support packaging for which there are a plethora of tutorials available online. There are many other important practices one should follow when crafting custom software, but I will leave those details to other works (such as Refs. \cite{wilson2014a, balaban2021, green2011}) as that can be an essay on its own. 

\begin{figure}
    \centering
    \includegraphics{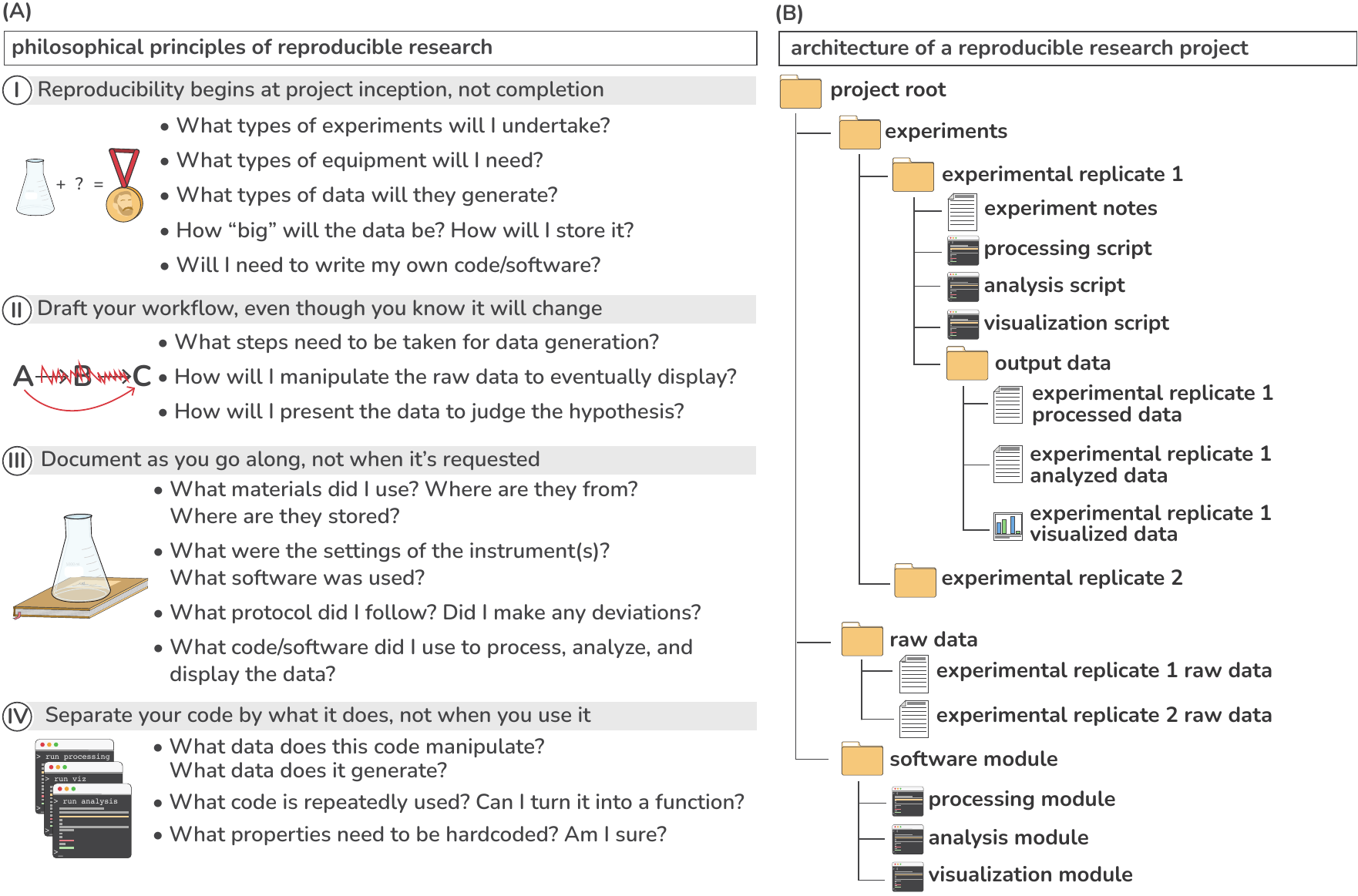}
    \caption{\textbf{Principles of reproducible research and a reproducible project architecture.}(A) The philosophical principles of reproducible research (I-IV) along with representative questions to ask as you perform your research. (B) A representative project architecture embodying the philosophical principles. A usable template of this architecture is available on GitHub at \href{https://github.com/gchure/reproducible_research}{\texttt{github.com/gchure/reproducible\_research}}}
    \label{fig:principles}
\end{figure}

\section*{Boilerplating the boring bits}
These principles and their associated reflexive questions [Figure \ref{fig:principles}(A)] have greatly helped me organize and perform my research reproducibly without sacrificing the adventurous spirit of actually doing the research. Rather than rigid rules, they act as prompts that force me to think about the little details that make the research work. 

Implementing them, however, can be less clear. I have found success in thinking of the principles as organizational guidelines for a file structure or directory tree, as is sketched out in Figure \ref{fig:principles}(B). At the beginning of every research project, I make a series of empty folders emulating this layout, which makes me think of what types of experiments I will perform and data I will generate (principle I). With that in place, I can outline the experimental protocols and procedures I will follow, as well as draft out any major code I will need to write (principle II). As the experiments proceed and data starts being generated, I write my experimental notes and fine-tune the scripts in real time (principle III) allowing me to assess my scientific hypotheses as they are tested. Outside of the specific experimental folders, I have my custom-written software package for the various steps of the scientific process (principle IV). 

I use this same structure in all of my research projects and always make the directory publicly accessible as GitHub repositories\footnote{See \href{http://github.com/rpgroup-pboc/mwc_mutants}{\texttt{github.com/rpgroup-pboc/mwc\_mutants}} (Ref. \cite{chure2019}) and \href{https://github.com/rpgroup-pboc/vdj_recombination}{\texttt{github.com/rpgroup-pboc/vdj\_recombination}} (Ref. \cite{hirokawa2020}) for some practical examples.}. In essence, this means that I publicly release \textit{the entirety of my lab notebook} upon submission of a manuscript. While this may not be possible for your research (always make sure you can do so legally!), I prefer to perform my science in the open. That way, if someone wants to see what code I ran for a particular experiment or what experiments didn't make it into the final cut of the manuscript, they are just a few mouse clicks away. 

Since all of my projects follow essentially the same structure, I have set up a template repository on GitHub (\href{https://github.com/gchure/reproducible_research}{\texttt{https://github.com/gchure/reproducible\_research}}) with extensive documentation and narration of the design logic. With a GitHub account, you can make a copy of this language agnostic template and immediately begin using it for your research. Over the past few years, I've had the joy of seeing other people use this structure in their projects, either wholesale or with reorganization to suit their particular needs. I invite the scientific community to use this template as a means to help lower the barrier for designing reproducible research projects by boilerplating the ``boring" bits.  

\section*{Reproducibility as a requirement for understanding}

\begin{figure}
    \centering
    \includegraphics{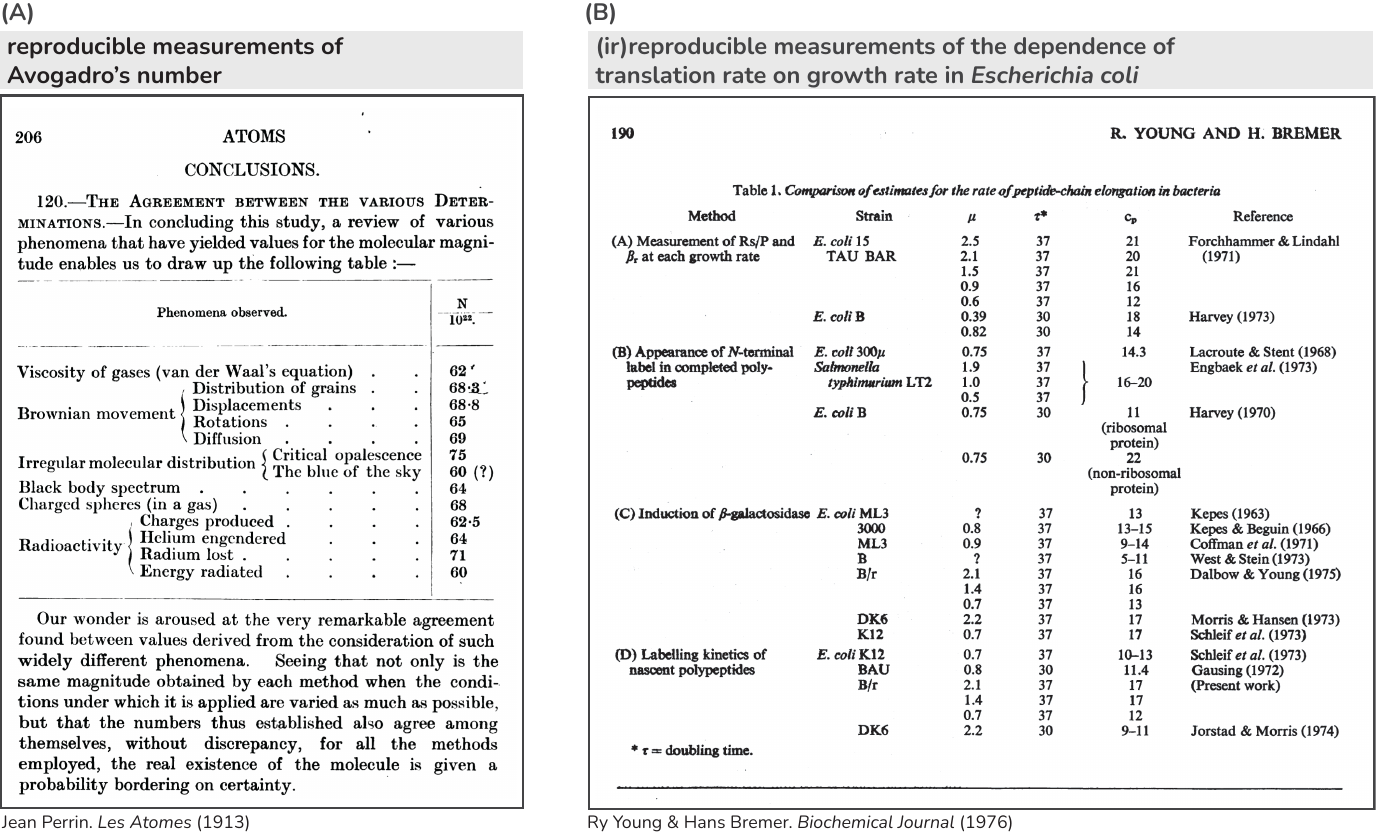}
    \caption{\textbf{Replicability and reproducibility of measurement in the physical and biological sciences.} (A) Jean Perrin's comparison of the inferred value of Avogadro's number from different researchers using wildly different methods. Table taken from Ref. \cite{perrin1913}. (B) Ry Young and Hans Bremer's collated measurements of the growth rate ($\mu$), doubling time ($\tau^*$), and peptide-chain elongation rate ($c_p$) in peptide bonds per second from different authors using different methods. Table taken from Ref. \cite{young1976}.}
    \label{fig:tables}
\end{figure}

It is simultaneously expected and surprising that scientific findings are reproduced with regularity and precision. Figure \ref{fig:tables} demonstrates two of my favorite examples of the implications of reproducibility, the repeated measurement of Avogadro's number (A) and measurement of the rate of protein translation in \textit{E. coli} (B). In both of these examples, completely different methods are used to measure the same property, with the degree of concordance between them determining whether understanding can be claimed. This is beautifully captured by the physicist Jean Perrin (who authored Figure \ref{fig:tables}(A) in Ref. \cite{perrin1913}), 
\begin{quote}
    \textit{``Our wonder is aroused at the very remarkable agreement found between values derived from the consideration of such widely different phenomena. Seeing that not only is the same magnitude obtained by each method when the conditions under which it is applied are varied as much as possible, but that the numbers thus established also agree among themselves, without discrepancy, for all the methods employed, the real existence of the molecule is given a probability bordering on certainty.''}
\end{quote}
Very rarely is one allowed to state anything with statistical certainty. The ``existence of the molecule,'' however, is one such case whose evidence is insurmountable and reproducible. 

In biology, we are often tempted to chalk up disagreement between methods, labs, or individual researchers to myriad factors that are out of our control. While studying living matter does present its own challenges, this should not be used as an excuse to lower the bar for what it means to truly understand. In Figure \ref{fig:tables}(B), Ry Young and Hans Bremer \cite{young1976} list the state-of-the-art measurements (at the time) of the speed at which \textit{E. coli} ribosomes can synthesize new proteins. Despite using the same organism, different studies produced different values for the rate, even when they were measured using the same method! This lack of agreement was in conflict with the prevailing view that this rate of translation was a fixed quantity---so why wasn't it reproducible? Rather than just chalking up the discrepancy to noise, Young and Bremer argued that this lack of reproducibility challenged this hypothesis, demanding more careful and direct experiments. Their identification of a growth-rate dependent translation rate has since been reproduced (for example in Refs. \cite{dai2016, wu2022}, and as a meta-analysis in Ref. \cite{chure2022}), providing a more concrete understanding of how this important phenomenon is regulated.

As scientists, we hold the precious responsibility of performing our research with objectivity and rigor. This is a feature of science that has held true throughout history regardless of the scientific questions at hand or the methods at play. As technological advancements transform our ability to examine the world around us, it also transforms the ways this responsibility is challenged. It is up to us to adapt the way we perform research to confront these novel threats to objectivity and rigor such that they remain sacrosanct.

\section*{Acknowledgements}
This work is the product of years of performing and troubleshooting research with a number of excellent collaborators and instructors. I thank Rachel Banks, Stephanie Barnes, Nathan Belliveau, Avi Flamholz, Soichi Hirokawa, Zofii  Kaczmarek, Mason Kamb, Heun Jin Lee, Ignacio Lopez-Gomez, Muir Morrison, Rob Phillips, Manuel Razo-Mejia, and Nicholas Sarai for their help in trying out different schemes of how to ensure reproducibility in our shared research efforts. I also thank Justin Bois at Caltech for his tremendous efforts in teaching me how to think of data in the biological sciences and how it should interact with software. I thank Suzy Beeler, Justin Bois, August Burton, Callie R. Chappell, RC (Rebecca Christensen), Jonas Cremer, Kian Faizi, Roshali de Silva, Soichi Hirokawa, Mathis Leblanc, Shaili Mathur, Rob Phillips, Rachel Porter, Manuel Razo-Mejia, Shyam Saladi, Richa Sharma, and Sophie Walton for critical feedback and suggested revisions to the manuscript. I also thank Rob Phillips for exposing me to the reproducible measurements of Avogadro's number as remarked upon by Jean Perrin. I am financially supported via the National Science Foundation Postdoctoral Fellowship in Biology under project 2010807.

\bibliography{library.bib}
\end{document}